\documentclass[pre,aps,preprint,showpacs]{revtex4}
\usepackage{revsymb}
\usepackage{graphicx,psfrag,nicefrac,subfigure}
\usepackage{amsmath}

\newcommand{\abs}[1]{\left\lvert#1\right\rvert}

\begin{document}
\author{Knud Zabrocki}
\affiliation{Fachbereich Physik,
Martin-Luther-Universit\"at,D-06099 Halle, Germany}
\email{zabrocki@physik.uni-halle.de}
\author{Svetlana Tatur}
\affiliation{Dept. of Physics, Ural State University, 620083 Ekaterinburg, Russia}
\email{svetlana.tatur@mail.ru}
\author{Steffen Trimper}
\affiliation{Fachbereich Physik,
Martin-Luther-Universit\"at,D-06099 Halle Germany}
\email{trimper@physik.uni-halle.de}
\author{Reinhard Mahnke}
\affiliation{Institut f\"ur Physik, Universit\"at Rostock, D--18051 Rostock, Germany} 
\email{reinhard.mahnke@uni-rostock.de}
\title{Relationship between a Non-Markovian Process and Fokker-Planck Equation}
\date{\today }

\begin{abstract}
We demonstrate the equivalence of a Non--Markovian evolution equation with a 
linear memory--coupling and a Fokker--Planck equation (FPE). In case the feedback term 
offers a direct and permanent coupling of the current probability density to an initial 
distribution, the corresponding FPE offers a non-trivial drift term depending itself on the 
diffusion parameter. As the consequence the deterministic part of the underlying Langevin 
equation is likewise determined by the noise strength of the stochastic part. This memory 
induced stochastic behavior is discussed for different initial distributions. The analytical 
calculations are supported by numerical results. 
\end{abstract}
\maketitle

\section{Introduction}

One century after Einstein's explanation of Brownian motion \cite{einstein1}, compare also \cite{einstein1b}, 
stochastic processes are ubiquitous in almost every branch of physics \cite{brownian} and beyond it like in 
economics \cite{black}, 
chemistry \cite{avr} and biology \cite{mur}. The notion "Brownian motion" not only stands for the movement 
of pollens suspended in water, 
but is a generic term for the Wiener process \cite{wiener}, see also the textbooks \cite{frank,gardiner,risken}.  
Two main approaches in describing stochastic processes has been developed, related to the Langevin and to the 
Fokker-Planck equation, respectively. The underlying basic for the stochastic description is the ability of 
a separation of time scales leading to an evolution equation with a deterministic part supplemented 
by a stochastic force. In this Langevin equation the probability of the stochastic force is in general an input. 
An alternative way is the treatment of the stochastic process by the Fokker-Planck equation (FPE) 
\cite{frank,gardiner,risken}, where a parabolic partial differential equation for a probability density 
is the basis of calculation. As already known both description are equivalent for special stochastic forces  
\cite{risken,gardiner,zwanzig}. In the last decades there is an increasing effort in generalization
of the Langevin equation respectively the Fokker-Planck equation by including memory effects 
\cite{ohira,henry,morgado,anomal}. In the framework of investigation of anomalous diffusion \cite{anomal} and 
fractional diffusion \cite{henry} the descriptive equations are non-Markovian (non-local in time). 
The combination of diffusion with feedback couplings has not only an influence  on the long time behavior, for
the stationary state, but on the dynamics in an intermediate time regime. Such a memory dominated behavior is 
well established in analyzing the freezing processes in undercooled liquids 
\cite{leutheusser,goetze1,goetze2,schulz1}, where the underlying mathematical representation is based
on a projector formalism proposed by Mori \cite{mori}. In \cite{trimper1} two of us studied a 
simple model, which still includes the dynamical features of evolution models as conservation of the 
relevant quantity $p (\vec{r},t)$, which could be interpreted as probability density for a particle and 
moreover, a time delayed feedback coupling. Due to this coupling the generic behavior may be changed 
by the additional delay effects and could lead to non-stationary steady state solutions.\\
The aim of the present paper is to show the equivalence between a special non-Markovian equation 
\cite{trimper1} and the standard FPE with a non-trivial drift term. 
Firstly we present the non-Markovian model and its main features which consists of the occurrence of a 
non-trivial stationary solution and the dependence of the solution on the initial distribution  
$p_0 (\vec{r})\equiv p(\vec{r},t=0)$. As the result the underlying FPE reveals a 
drift term which is itself determined by the strength of the stochastic force. The corresponding potential 
are discussed in detail. The analytical results are supported by numerical simulations. 

\section{The Non-Markovian Fokker-Planck Equation}

In media with a spatial--temporal accumulation process transport 
phenomena should be described by a stochastic approach~\cite{MKL} based on 
probabilities. The time evolution of the probability density could
depend on the history of the sample to which it belongs, i.~e. the
changing rate of the probability should be influenced by the changing
rate in the past and so the evolution equation of the probability has to
be supplemented by memory terms. A recent overview is given 
in~\cite{frank}. The obvious modification of the FPE by memory effects is 
to replace the conventional equation by \cite{trimper1}
\begin{equation}
\partial_{t} p (\vec{r},t) = \mathcal{M}(\vec{r},t;p,\nabla p) +
\int\limits_{0}^{t}dt' \int\limits_{-\infty}^{\infty} d^{d}r' 
\mathcal{K}(\vec{r}-\vec{r}',t-t';p,\nabla p)\, 
\mathcal{L}(\vec{r}',t';p,\nabla p)\,.
\label{eq:01}
\end{equation}
This equation is of convolution type and consists of two competing parts
standing for processes on different timescales. The first part
manifested by the operator $\mathcal{M}$ characterizes the
instantaneous and local process, whereas the second part with the
operators $\mathcal{K}$ and $\mathcal{L}$ represent the delayed
processes, the memory. In general all the operators may be non-linear in
$p(\vec{r},t)$ and $\nabla p(\vec{r},t)$. Physically it means that the time scale of the memory is 
determined by the relevant probability $p$ itself. The specification of the operators 
has to be according to the physical situation, which one deals with. However the main feature of 
the probability density $p(\vec{r},t)$ is its conservation in time:
\begin{equation}
\frac{dP(t)}{dt}=\frac{d}{dt}\int\limits_{-\infty}^{\infty} d^d r \, p(\vec{r},t)
= 0 \, .
\label{eq:02}
\end{equation}
To preserve $p$ the instantaneous term $\mathcal{M}$ has to be related
to a probability current, e.~g., $\mathcal{M}\propto \nabla\cdot
\vec{j}$. For an arbitrary polynomial kernel $\hat{K}(\vec r, t) $ the conservation law 
(\ref{eq:02}) is not fulfilled in general. Using Laplace transformation one can show directly, compare 
for details~\cite{trimper1}, that the choice $\mathcal{L}\equiv - \, \partial_t p(\vec{r},t)$ guarantees 
conservation. Thus our starting equation is written in the form 
\begin{equation}
\partial_t p(\vec r,t) = D \, \nabla^2 p(\vec r,t)-
\int\limits_{0}^{t}dt'\int\limits_{-\infty}^{\infty} d^{d}r' K(\vec r -\vec r ',t-t')\, 
\partial_{t'} p(\vec r ',t')
\label{evo}
\end{equation}
as a special realization of Eq.~(\ref{eq:01}). This Fokker-Planck equation relates $p(\vec r,t)$
to $p(\vec r,t')$ with $0<t'<t$ unlike to a conventional one, where 
the evolution is only dependent on the probability at present time.
Moreover \eqref{evo} offers a coupling between $\partial_t p(\vec r,t)$
and $\partial_{t'} p(\vec{r},t')$. The mixing of time scales leads to a
substantial modification of the long time limit. To demonstrate the modification let us 
consider a very simple choice by a strictly spatial local, but time independent
kernel
\begin{equation}
K(\vec r,t)=\mu\, \delta (\vec{r}) \, ,
\end{equation}
where parameter $\mu>0$ characterizes the strength of the memory. By this choice
the spatial and temporal variables are decoupled. Inserting the kernel
in \eqref{evo} one gets
\begin{equation}\label{evo2}
\partial_{t} p (\vec{r},t)=D\,\nabla^2 p(\vec{r},t) - 
\mu\, \left[p(\vec{r},t)-p_0 (\vec{r}) \right]  \quad\text{with}\quad
p_0 (\vec{r})\equiv p(\vec{r}, t=0) \, .
\end{equation}
The time independence of the kernel means that all times $t'$ $(0<t'<t)$
in the past have the same weight and so there is a very strong 
memory with a direct coupling of the instantaneous value to the initial
value. The memory induced feedback to the initial value appears as a
driving force. Without this coupling one can interpret the equation as a
description of a particle, which performs a diffusive motion, where the
probability density $p (\vec{r},t)$ decays on a time   scale $\mu^{-1}$\,.
As \eqref{evo2} is a linear equation and so the solution of it could be 
found analytically for arbitrary initial conditions
\begin{equation}\label{evo3}
p(\vec{r},t) = \text{e}^{-\mu\, t} 
\int\limits_{-\infty}^{\infty} d^{d} r'\,  p_0 (\vec{r}\,')
\left[ G(\vec{r}-\vec{r}\,',t)\,  +\mu\,  
\int\limits_{0}^{t} dt'  \,G(\vec{r}-\vec{r}\,',t-t')\, \text{e}^{\mu \, t'} \right] 
\; ,
\end{equation}
where $G(\vec{r},t)$ is the Green's function of the conventional
diffusion equation 
$$
G(\vec r, t) = \frac{\Theta (t)}{(4 \pi D t)^{d/2}} \exp(-\vec r\,^2 /4 D t)\,.
$$ 
From the general solution, some properties could be
followed easily such as if the initial distribution is non-negative $p_0
(\vec{r})$, so the $p(\vec{r},t)$ does provided $\mu >0$. The second
moment $s(t)$ could be calculated
\begin{equation}
s(t)\equiv \int\limits_{-\infty}^{\infty}  \vec{r}^2\, p(\vec{r},t) \,d^d r =
\frac{2\, d\, D\, (1-\text{e}^{-\mu\, t})}{\mu}
\int\limits_{-\infty}^{\infty} p_0 (\vec{r})\, d^d r + 
\int\limits_{-\infty}^{\infty} \vec{r}\,^2 \, p_0 (\vec{r})\, d^d r \, .
\end{equation}
Notice that for the limit of vanishing memory $\mu \to 0$ one can
easily verify that the last equation shows conventional diffusive behavior.
The selection of the initial distribution is the essential point in our
model and so three example are given to illustrate the solution of
\eqref{evo2}. Without lack of generality we concentrate our calculation
on the one-dimensional case. It can be generalized to higher dimensions.

\section{Results}
Obviously the results for a system with memory should be sensitive with respect to 
the initial distribution or at least from configurations in the past. Therefore, 
we study different realizations for $p_0(\vec r)$ separately. 
\subsection{Stationary solution}

The first example is the delta--starting distribution 
$p_0 (x)= p_0\, \delta (x)$. Substituting this in 
\eqref{evo3} the following solution is calculated  
\begin{align}
  p(x,t)&=\frac{p_0}{\sqrt{4\, \pi \, D \, t}}\, \text{e}^{-\left(\mu\, t +\frac{x^2}{4\,D\, t}\right)}+\frac{p_0\, \kappa}{4}\,\left[f_{+}(x;D, \mu )+f_{-}(x;D, \mu )\right]   \\
& \nonumber \\
  f_{\pm}&=\text{e}^{\pm\kappa\, x}\,\left[\text{erf}\left(\frac{\pm x}{\sqrt{4\, D\, t}}
  +\sqrt{\mu\, t}\right)-\text{sgn}(\pm x)\right] \, ,
\end{align} 
where $\mbox{\text{erf}(x)}$ is the error function \cite{abram} and  $\kappa =\sqrt{\mu/D}$. The first part is the 
solution of the  homogeneous equation showing temporal decay with time constant $\mu^{-1}$. In the
long time limit the system shows a non-trivial stationary solution
\begin{equation}
\lim\limits_{t\to\infty} p(x,t) \equiv p_s (x) =\frac{p_0\, \kappa}{2}\, 
\text{e}^{-\kappa\, \abs{x}} \, .
\end{equation}
Such a stationary solution is due to the permanent coupling to the
initial distribution and the greater $\mu>0$ the stronger is this effect
and more pronounced are the deviations from the pure diffusive behavior
($\mu =0$). The generalization to higher dimensions $0 < d < 5 $ and arbitrary initial condition 
can be directly calculated. It results in   
\begin{equation}\label{evo4}
p_s (\vec r)=\frac{\kappa^{\frac{d+2}{2}}}{\left(2\, \pi\right)^{\frac{d}{2}}}
\int d^d r' \, \frac{p_0 (\vec r')}{\abs{\vec r -\vec r'}^{\frac{d-2}{2}}}\, 
\text{K}_{\frac{d-2}{2}}\left(\kappa\,\abs{\vec r -\vec r'}\right)\, ,
\end{equation}
with $\kappa^2 = \mu/D$ and $K_{\nu}(x)$ is a modified Bessel function ~\cite{abram}.

Using the last result in Eq.~(\ref{evo4}) we get for the Gaussian distribution 
$p_0 (x)=p_0\, \text{e}^{-\lambda\,x^2}$, the following result
\begin{align}
p_s (x) &= \frac{p_0\, \kappa}{4}\,\sqrt{\frac{\pi}{\lambda}}\, \text{e}^{\beta^2}\, \left[g_{+}(x;\beta ,\lambda , \kappa )+g_{-}(x;\beta ,\lambda , \kappa )\right]\\
\intertext{with}
g_{\pm}(x;\beta ,\lambda , \kappa )&=\text{e}^{\pm \kappa \, x}\, \text{erfc}\left(\beta\pm x\sqrt{\lambda}\right)\quad
\text{and}\quad
\beta =\sqrt{\frac{\mu}{4\, \lambda \, D}}=\frac{\kappa}{2\, \sqrt{\lambda}} \, .
\end{align}
Here erfc$(x)$ is the complementary error function \cite{abram}. In case of an exponential 
decreasing initial distribution 
$p_0 (x)=p_0\, \text{e}^{-\lambda\, \abs{x}}$ the calculation leads to
\begin{equation}
p_{s} (\abs{x})=\begin{cases}
p_0\, \frac{\lambda\, \kappa^2}{\kappa^2 -\lambda^2}\, \left[\frac{\text{e}^{-\lambda\, \abs{x}}}{\lambda}-\frac{\text{e}^{-\kappa\, \abs{x}}}{\kappa}\right]&\quad\text{for}\quad \lambda\not= \kappa\\
&\\
p_0 \, \frac{1+\kappa\, \abs{x}}{2}\, \text{e}^{-\kappa\, \abs{x}}&\quad\text{for}\quad \lambda=\kappa \quad .
\end{cases}
\end{equation}
Let us note that the stationary distribution depends on all cases on the initial conditions. 

\subsection{Relationship to Fokker-Planck Equation }
\begin{figure}
\subfigure[$\delta (x)$]{\includegraphics[height=3.3cm,width=3.5cm]{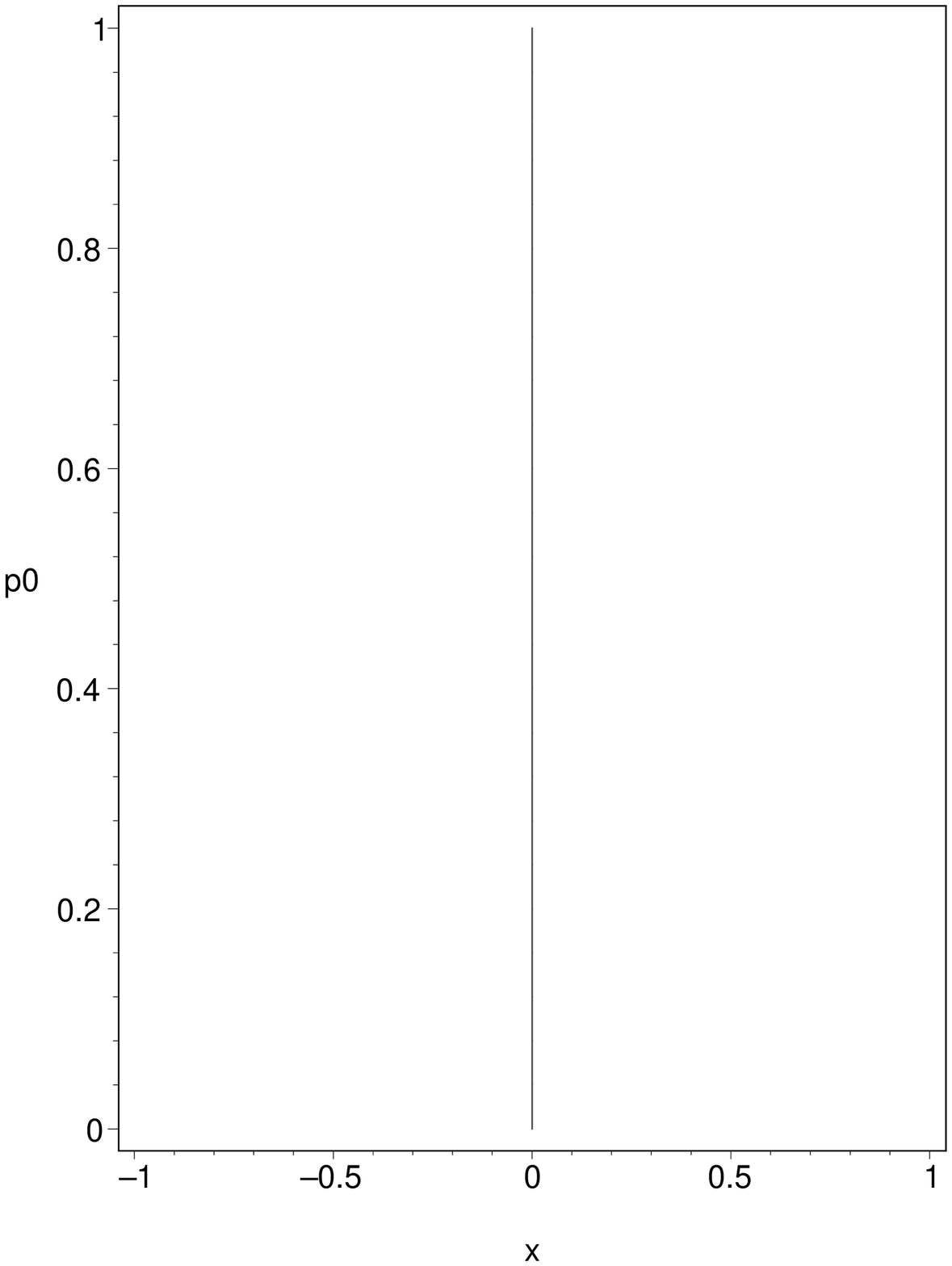}}
\hspace{0.1cm}
\subfigure[$\exp\left(-x^2 \right)$]{\includegraphics[height=3.3cm,width=3.5cm]{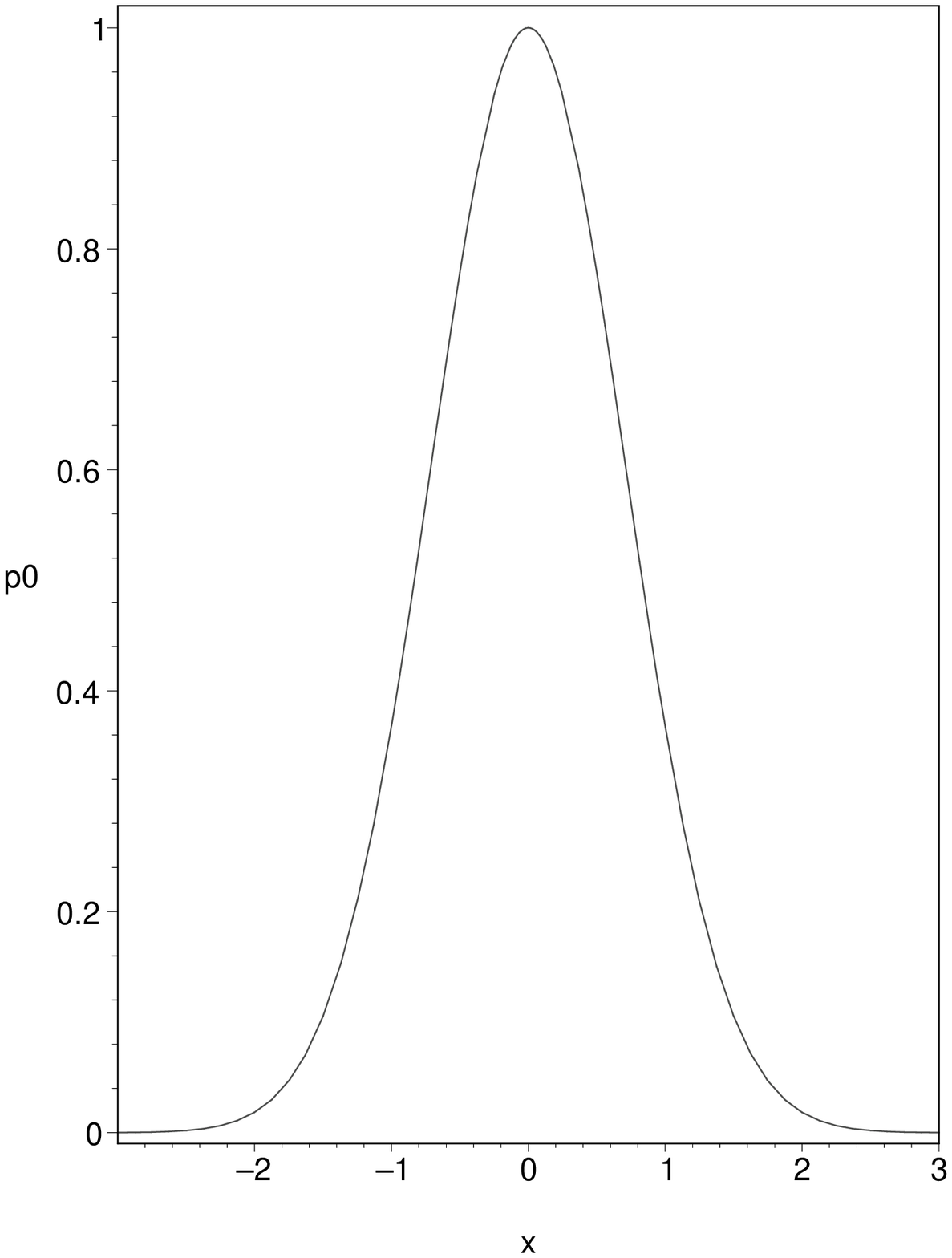}}
\hspace{0.1cm}
\subfigure[$\exp\left(-\abs{x}\right)$]{\includegraphics[height=3.3cm,width=3.5cm]{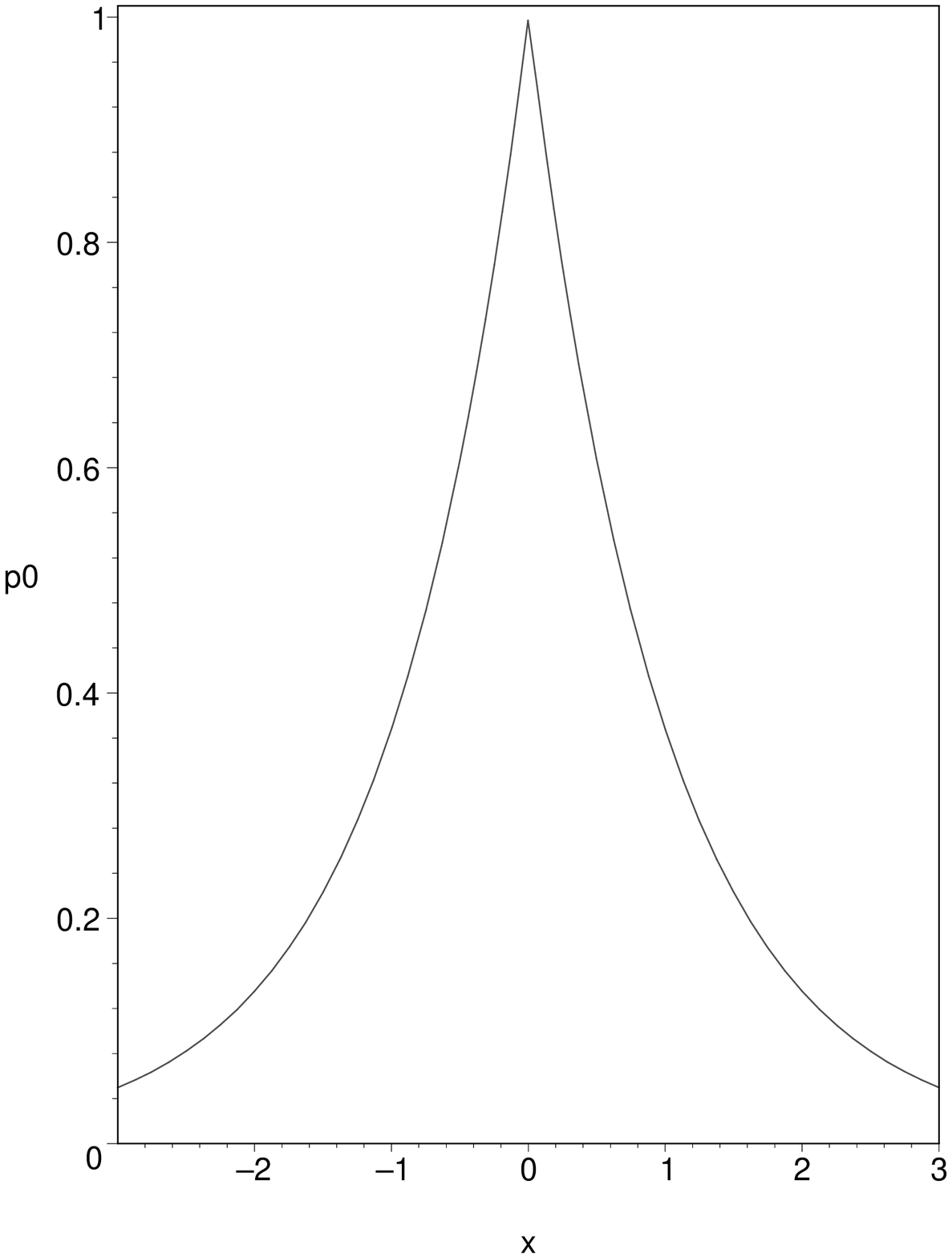}}
\caption{
\label{fig:1}
Starting distribution $p_0 (x)$, $p_0 =1$ and $\lambda =1$.}
\end{figure}

\begin{figure}
\subfigure[$\delta (x)$]{\includegraphics[height=3.3cm,width=3.5cm]{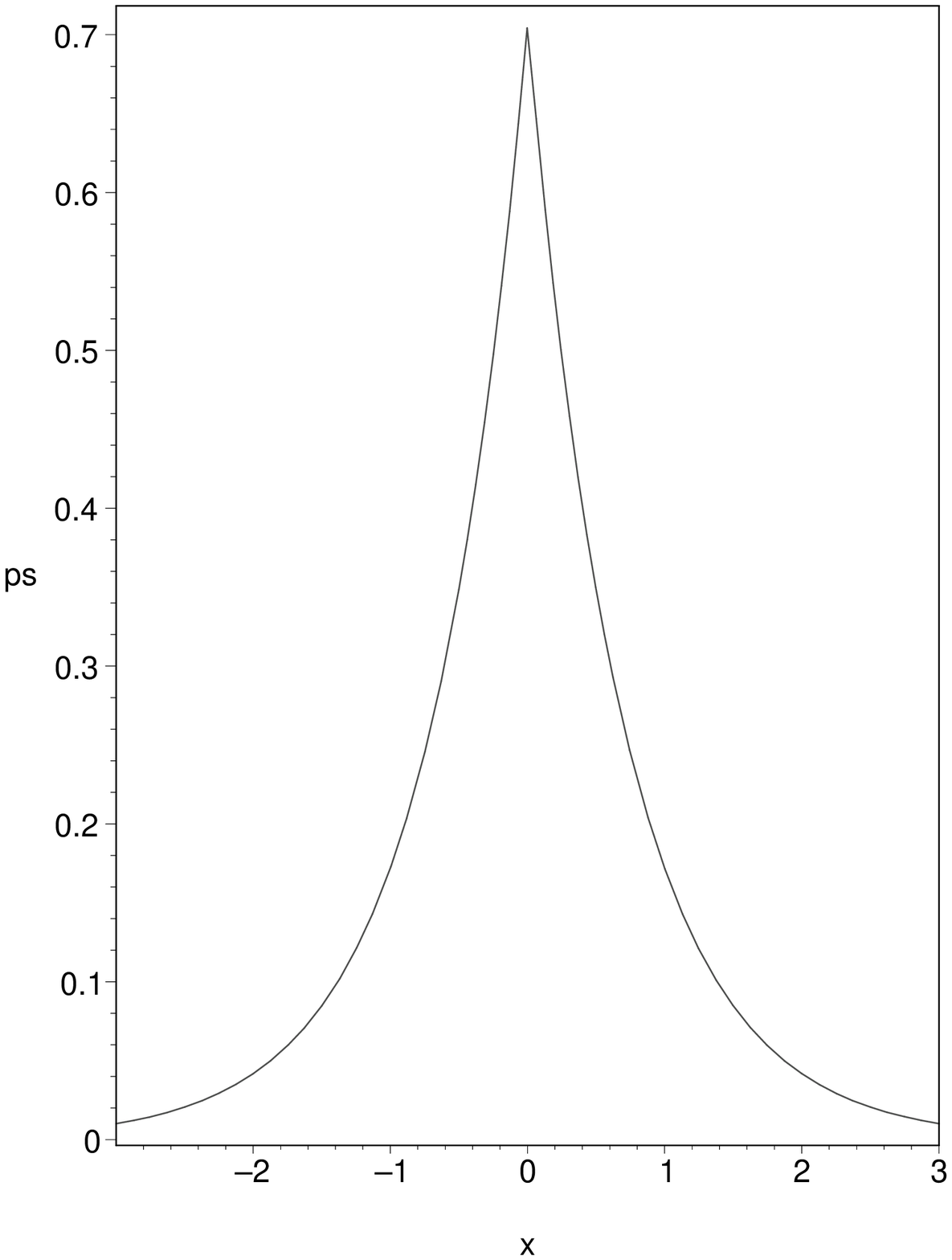}}
\hspace{0.1cm}
\subfigure[$\exp\left(-x^2 \right)$]{\includegraphics[height=3.3cm,width=3.5cm]{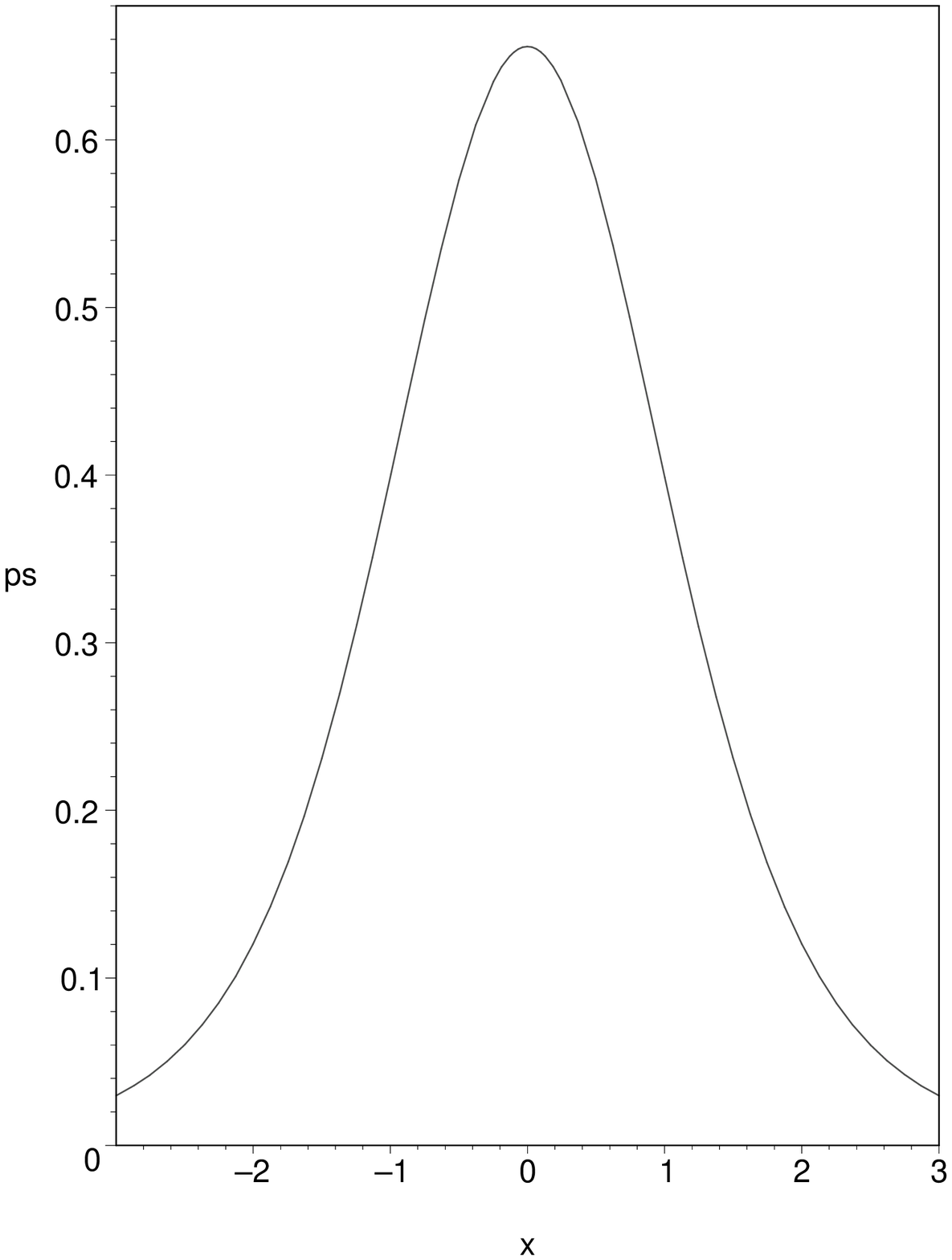}}
\hspace{0.1cm}
\subfigure[$\exp\left(-\abs{x}\right)$]{\includegraphics[height=3.3cm,width=3.5cm]{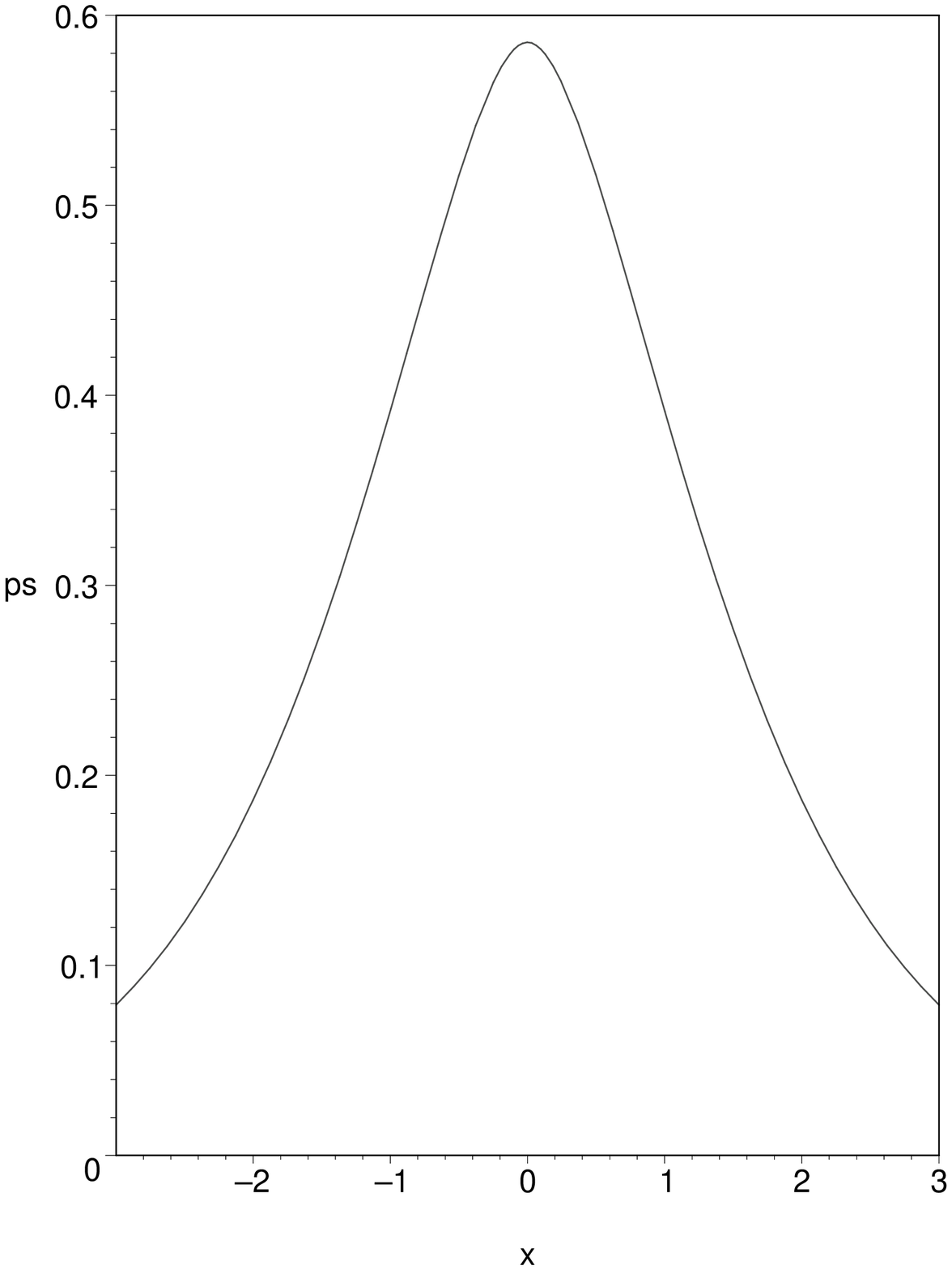}}
\caption{
\label{fig:2}
Stationary state $p_s (x)$, $p_0 =1$, $\lambda =1$, $D=1$ and $\mu =2$ (units arbitrary).} 
\end{figure}

\begin{figure}
\subfigure[$\delta (x)$]{\includegraphics[height=3.3cm,width=3.5cm]{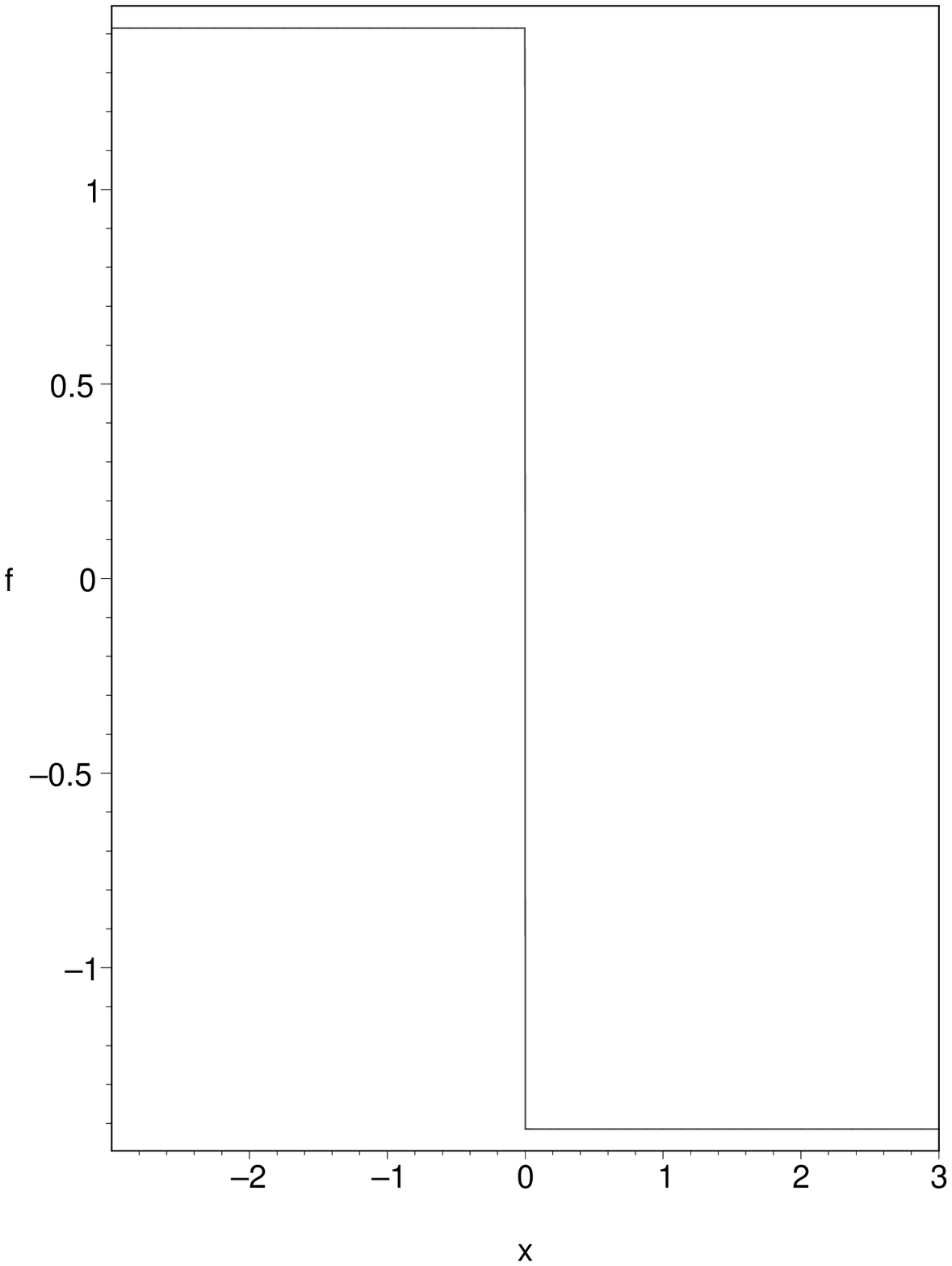}}
\hspace{0.1cm}
\subfigure[$\exp\left(-x^2 \right)$]{\includegraphics[height=3.3cm,width=3.5cm]{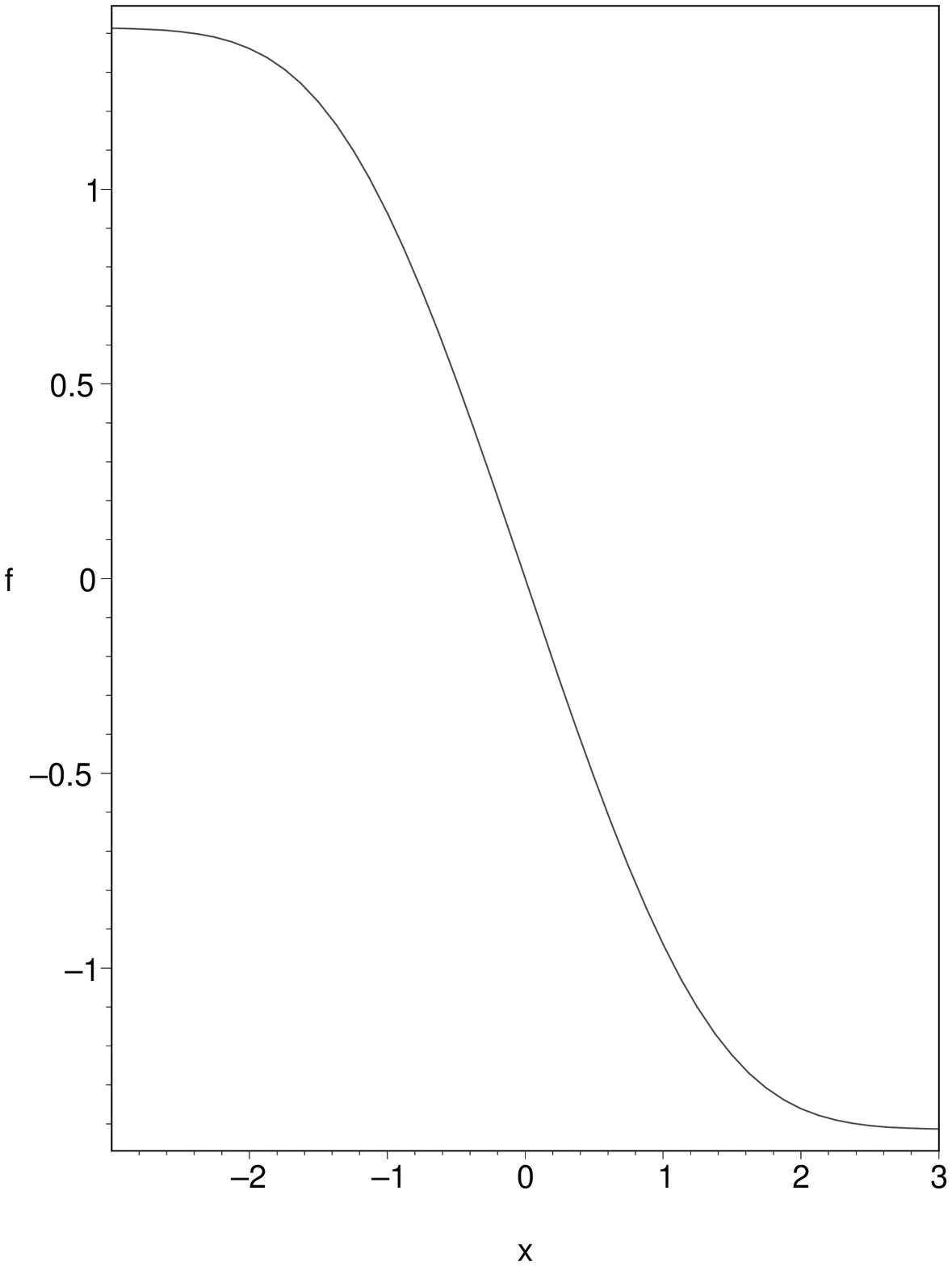}}
\hspace{0.1cm}
\subfigure[$\exp\left(-\abs{x}\right)$]{\includegraphics[height=3.3cm,width=3.5cm]{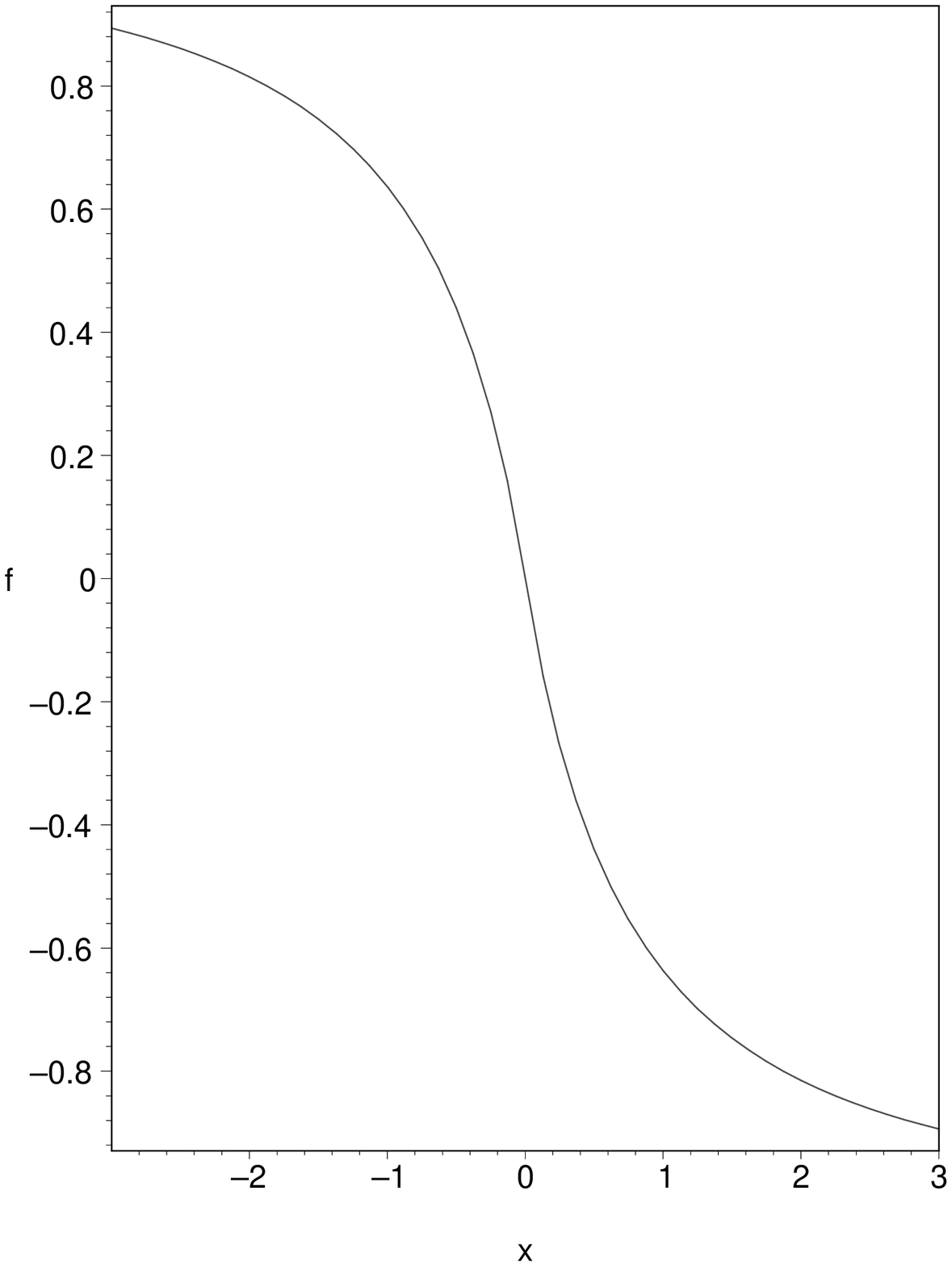}}
\caption{
\label{fig:3}
Drift coefficient $f(x)$, $p_0 =1$, $\lambda =1$, $D=1$ and $\mu =2$ (units arbitrary).} 
\end{figure}

\begin{figure}
\subfigure[$\delta (x)$]{\includegraphics[height=3.3cm,width=3.5cm]{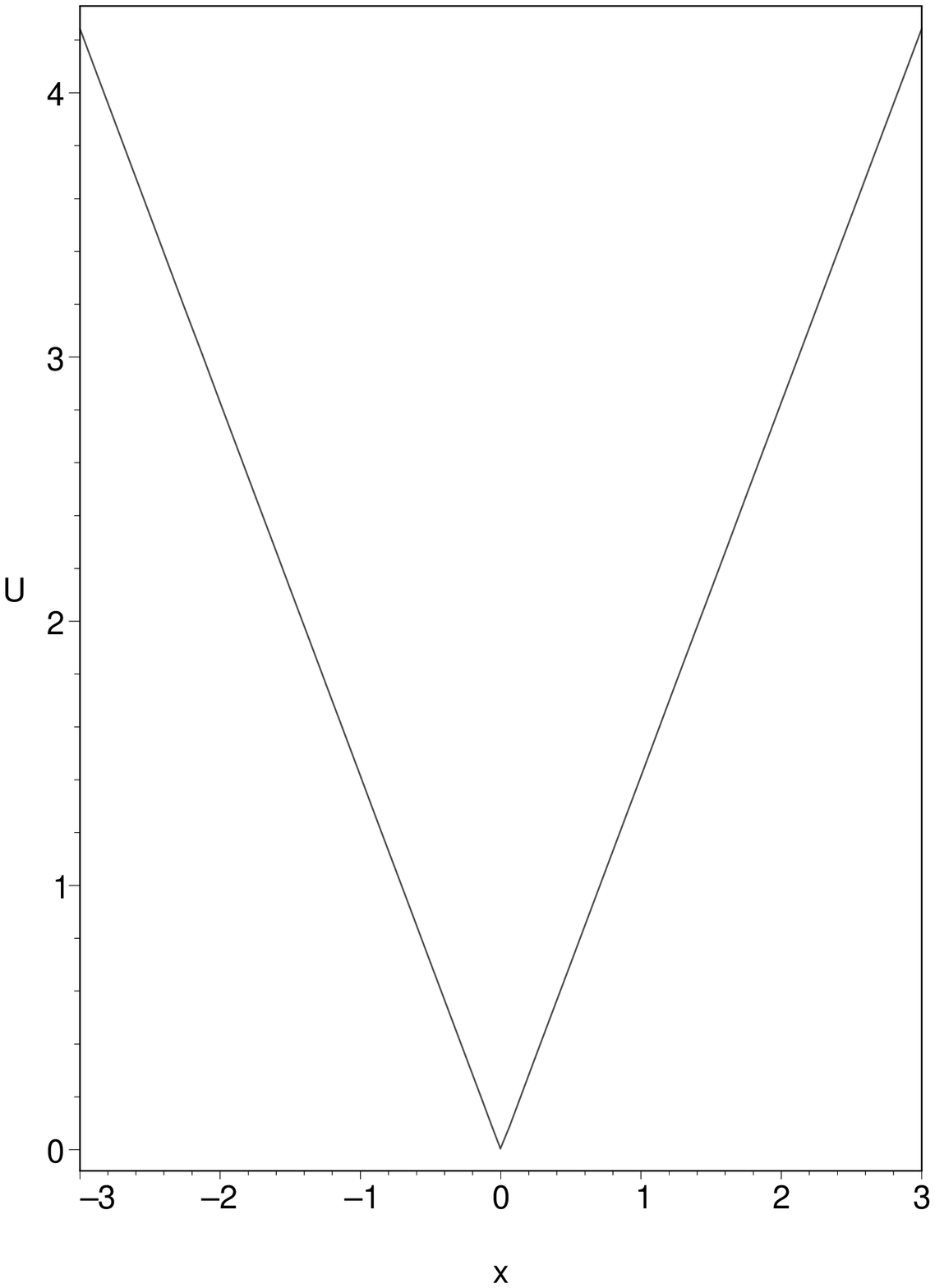}}
\hspace{0.1cm}
\subfigure[$\exp\left(-x^2 \right)$]{\includegraphics[height=3.3cm,width=3.5cm]{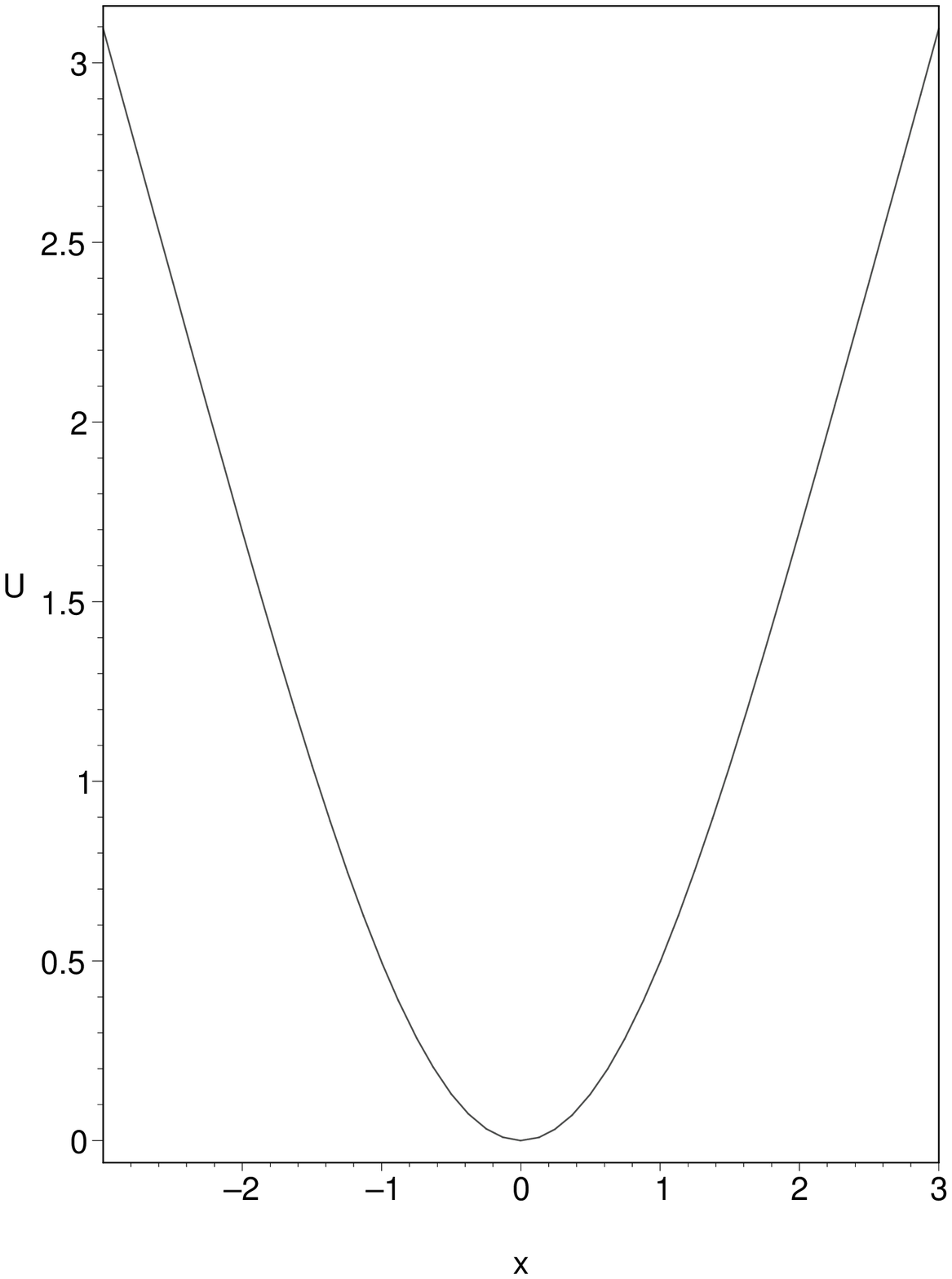}}
\hspace{0.1cm}
\subfigure[$\exp\left(-\abs{x}\right)$]{\includegraphics[height=3.3cm,width=3.5cm]{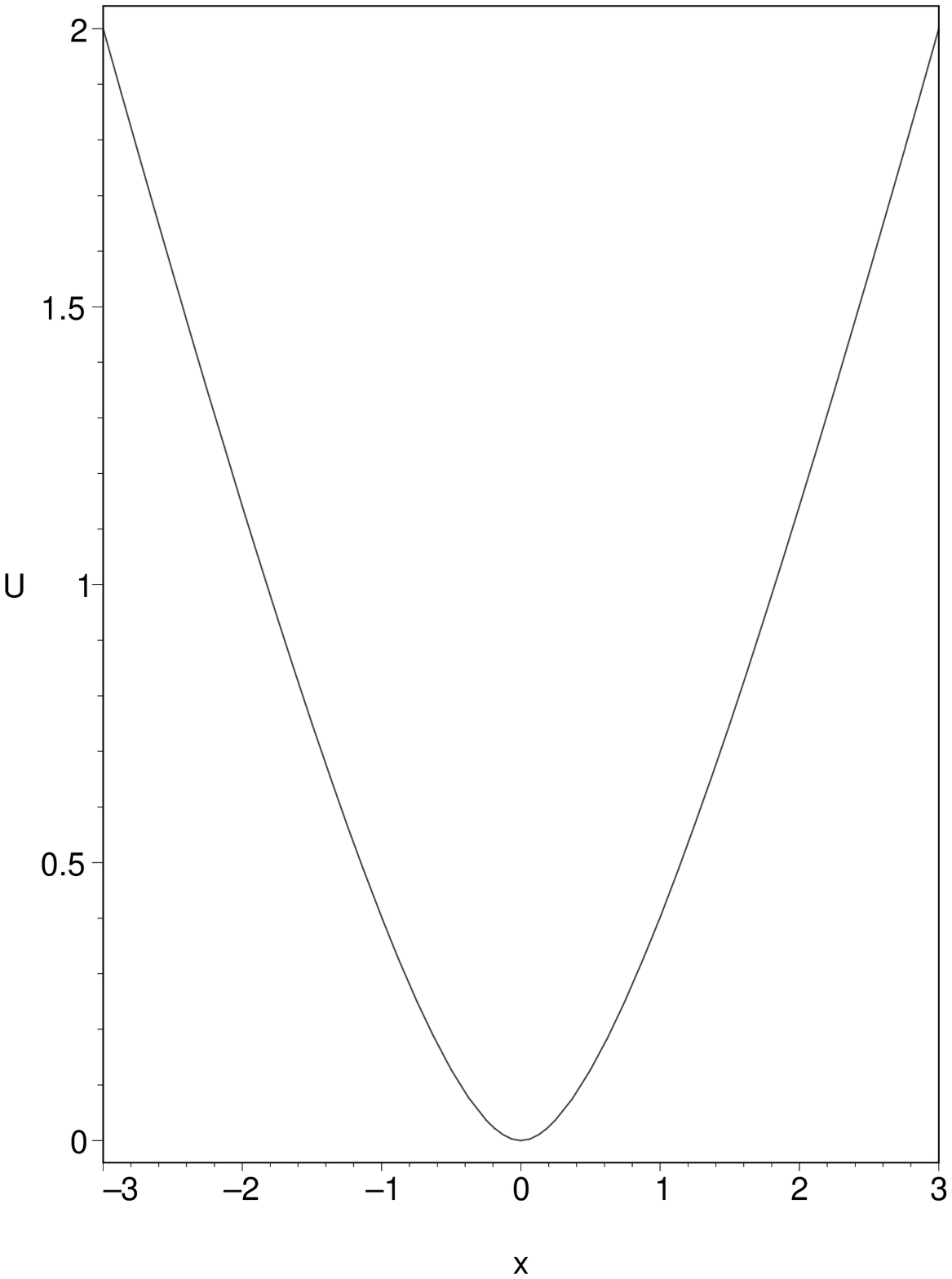}}
\caption{
\label{fig:4}
Potential $U(x)$, $p_0 =1$, $\lambda =1$, $D=1$ and $\mu =2$ (units arbitrary).} 
\end{figure}

The conventional form of the FPE including an external force, has in the one-dimensional case the following form
\begin{equation}\label{evo5}
\frac{\partial p (x,t)}{\partial t}=D\, \frac{\partial^2 p(x,t)}{\partial x^2} - 
\frac{\partial}{\partial x}\left[f(x)\, p(x,t)\right]\; ,
\end{equation}
where $D$ is the diffusion coefficient, supposed to be constant here, and $f(x)$ is the drift 
force, for which  $f(x)= -\, dU(x)/dx$ with $U(x)$ as
corresponding potential. On the one hand, the diffusion coefficient $D$
measures the intensity of the noise and represents the stochastic part
of motion, whereas the drift coefficient $f(x)$ corresponds to the force
experienced by the system and so it describes the deterministic part
of motion. In this subsection we calculate the force $f(x)$ which 
corresponds to the
potential $U(x)$ for the different starting distributions $p_0(x)$,
in such a way that both Fokker-Planck equations \eqref{evo2} and
\eqref{evo5} are equivalent. To do this deterministic parts of both
equations are compared in the long time limit, in the stationary state.
Provided $p_s (x) \not= 0$ the formal solution is found by integration
\begin{equation}\label{solution}
f(x)=\frac{\mu\int\limits_{-\infty}^{x}\left[p_s (\xi )-p_0 (\xi )\right]\, d\xi}{p_s (x)}
+\frac{C}{p_s (x)} \; ,
\end{equation}
where $C$ is an integration constant, which one could set to zero. To
show the equivalence mathematically correct, one has to do this
comparison in the following way. First take an arbitrary function $h(x)$
with bounded support, then integrate the product of $h$ and the
deterministic part of \eqref{evo2} resp. \eqref{evo5} over the complete
real line, and finally compare the results of these integrations. If
both integrations are equal, then the functions are equal. 
In case of arbitrary dimension the $\alpha $- component of the drift force is given by 
$$
f_{\alpha }  = \frac{\mu \int_{-\infty }^{x_{\alpha }} \left[p_s (\xi )-p_0 (\xi )\right]\, d\xi}{p_s (x)}\,.
$$
Here the integration constant is assumed to be zero as in the one dimensional case. 
Finally we present the results 
for three different initial distributions,
depicted in Fig.~\ref{fig:1}, the corresponding stationary solutions is shown in
Fig.~\ref{fig:2}, the drift term and the corresponding potential in 
Fig.~\ref{fig:3} and
Fig.~\ref{fig:4}, respectively. For the delta-like starting distribution the drift
term can be calculated to 
\begin{equation}
f(x)=-\sqrt{\mu\,D}\, \text{sign}(x)
\end{equation} 
and so one can verify the following potential
\begin{equation}
U(x)=\sqrt{\mu\,D}\, \abs{x} \; .
\end{equation}
In case of a Gaussian initial distribution an analogue calculation leads to
\begin{equation}
f(x) = \sqrt{\mu\,D}\,
\frac{g_{+}(x;\beta ,\lambda ,\kappa )-g_{-}(x;\beta ,\lambda ,\kappa)}
{g_{+}(x;\beta ,\lambda ,\kappa )+g_{-}(x;\beta ,\lambda ,\kappa )} \; ,
\end{equation}
whereas an exponential starting distributions yields to the drift term
\begin{equation}
f(x)=-\sqrt{\mu\, D}\,\text{sign}(x)\, \lambda\,
\begin{cases}
 \frac{\text{e}^{-\lambda\, \abs{x}}-\text{e}^{-\kappa\, \abs{x}}}
{\kappa\,\text{e}^{-\lambda\, \abs{x}}-\lambda\, \text{e}^{-\kappa\, \abs{x}}} & \quad\text{for}\quad \kappa \not= \lambda\\
\frac{\abs{x}}{1+\lambda\,\abs{x}}&\quad\text{for}\quad \kappa =\lambda
\end{cases}
\; .
\end{equation}
The underlying potentials can be obtain after an integration. The
results are shown in Fig.~4. It should be noticed that the potential is likewise determined by the 
diffusion constant $D$ which is a measure of the stochastic force. This point will be discussed in 
the next section.

\subsection{Deterministic motion in the potential}
As discussed before the speciality of the present approach consists of including the diffusive parameter 
$D$ likewise into the force term $f(x)$ of the FPE. Insofar it seems to be reasonable to study  
the corresponding Langevin equation, especially in the small noise limit. The trajectories  
obey the equation 
\begin{equation}
\frac{dx(t)}{dt}= f[x(t)] +\xi (t) \quad \rm{with} \quad \langle \xi (t)\, \xi (t') \rangle  = 2 D\, \delta (t - t')\,.
\label{lang}
\end{equation}
Firstly let us solve the deterministic equation of motion under the influence of the different potentials 
depicted in Fig.~\ref{fig:4}. All potentials offer their (global) minimum at $x\equiv 0$, so this is the 
fixed point of the equation  
\begin{equation*}
\frac{dx(t)}{dt}= f[x(t)]\quad .
\end{equation*} 
For the Delta-starting distribution one can calculate the deterministic trajectory exactly
\begin{equation*}
x(t;\mu ,D,x_0 )=\Theta \left(t-\frac{\abs{x_0}}{\sqrt{\mu\, D}}\right)\, \left[x_0 - \text{sign}(x_0 )\, \sqrt{\mu\, D} \, t\right]\quad .
\end{equation*}
There is a linear decrease (increase) of $x$ to the fixed point 
$x\equiv 0$ in the time interval $0 \leq t \leq \nicefrac{\abs{x_0}}{\sqrt{\mu\, D}}$. 
After reaching the fixed point the particle stays there if no noise is present. 

\begin{figure}
\centering
\includegraphics[height=5.0cm,width=5.0cm]{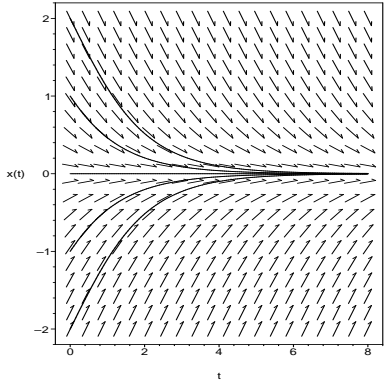}
\caption{
\label{fig:6}
Numerical calculated trajectory for $p_0 (x) =\exp \left(-x^2\right)$, $D=1$, $\mu =1$ and $x_0 =-2;-1;0;1;2$}
\end{figure}

\begin{figure}
\centering
\includegraphics[height=5.0cm,width=5.0cm]{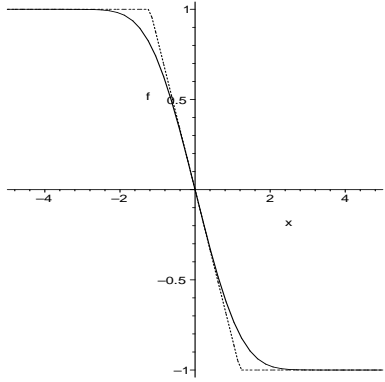}
\caption{\label{fig:7} Comparison between the drift term $f(x)$ (solid) and the piecewise approximative 
drift term $f_a (x)$ (dashdotted) for $\lambda =1,\mu =1, D=1$}
\end{figure}

For the Gaussian distribution the drift term is highly nonlinear and the deterministic equation is not 
solvable exactly. In Fig.~\ref{fig:6} the direction field of the solution of the deterministic motion is 
illustrated for different starting values, where the numerical calculation is done with the Runge-Kutta
$4^{th}$ order algorithm with a step length $h=0.05$.
As seen from Fig.~\ref{fig:4} (b) the potential exhibits two different regimes, namely a parabolic one if 
$\abs{x}\ll 1$ and linear regime for $\abs{x} \gg 1$. Assuming a piecewise approximation for the drift 
term as shown in Fig.~\ref{fig:7} the trajectories could be calculated exactly. Both pieces are merged together at 
the new parameter $x_0$
\begin{equation*}
x_a (t) =
\begin{cases}
\begin{cases}
x_0 +\sqrt{\mu\, D}\, t&\quad\text{for}\quad t\le t_1\\
\sqrt{\mu\, D}{s}\,\text{e}^{s\, (t-t_1 )}&\quad\text{for}\quad t> t_1\\
\end{cases}
&\quad\text{if}\quad x_0 <\frac{\sqrt{\mu\, D}}{s} \\
x_0\, \text{e}^{s\, t}&\quad\text{if}\quad \frac{\sqrt{\mu\, D}}{s}<x_0 <-\frac{\sqrt{\mu\, D}}{s} \\
\begin{cases}
x_0 -\sqrt{\mu\, D}\, t&\quad\text{for}\quad t\le t_2\\
-\sqrt{\mu\, D}{s}\,\text{e}^{s\, (t-t_2 )}&\quad\text{for}\quad t> t_1\\
\end{cases}
&\quad\text{if}\quad x_0 >-\frac{\sqrt{\mu\, D}}{s} \quad ,
\end{cases}
\end{equation*}
where  $t_1 =s^{-1}-\nicefrac{x_0}{\sqrt{\mu\, D}}$ and $t_2 =s^{-1}+\nicefrac{x_0}{\sqrt{\mu\, D}}$. 
In the outer area the decrease of $x(t)$ is linear, whereas after the cross-over it is exponential 
in the inner area. Because the piecewise approximation is quite rough we have studied a better one 
by replacing the drift term by 
\begin{equation}\label{tanh}
f(x)=\sqrt{\mu\, D}\tanh \left(\frac{s}{\sqrt{\mu\, D}}\, x\right) \quad ,
\end{equation}
where both regimes are matched, the long distance regime $x\to\pm\infty$ as well as the short 
distance regime $x \to 0$. 
\begin{figure}
\subfigure[$f(x)$ and $f_a (x)$  \eqref{tanh}]{\includegraphics[height=3.3cm,width=3.5cm]{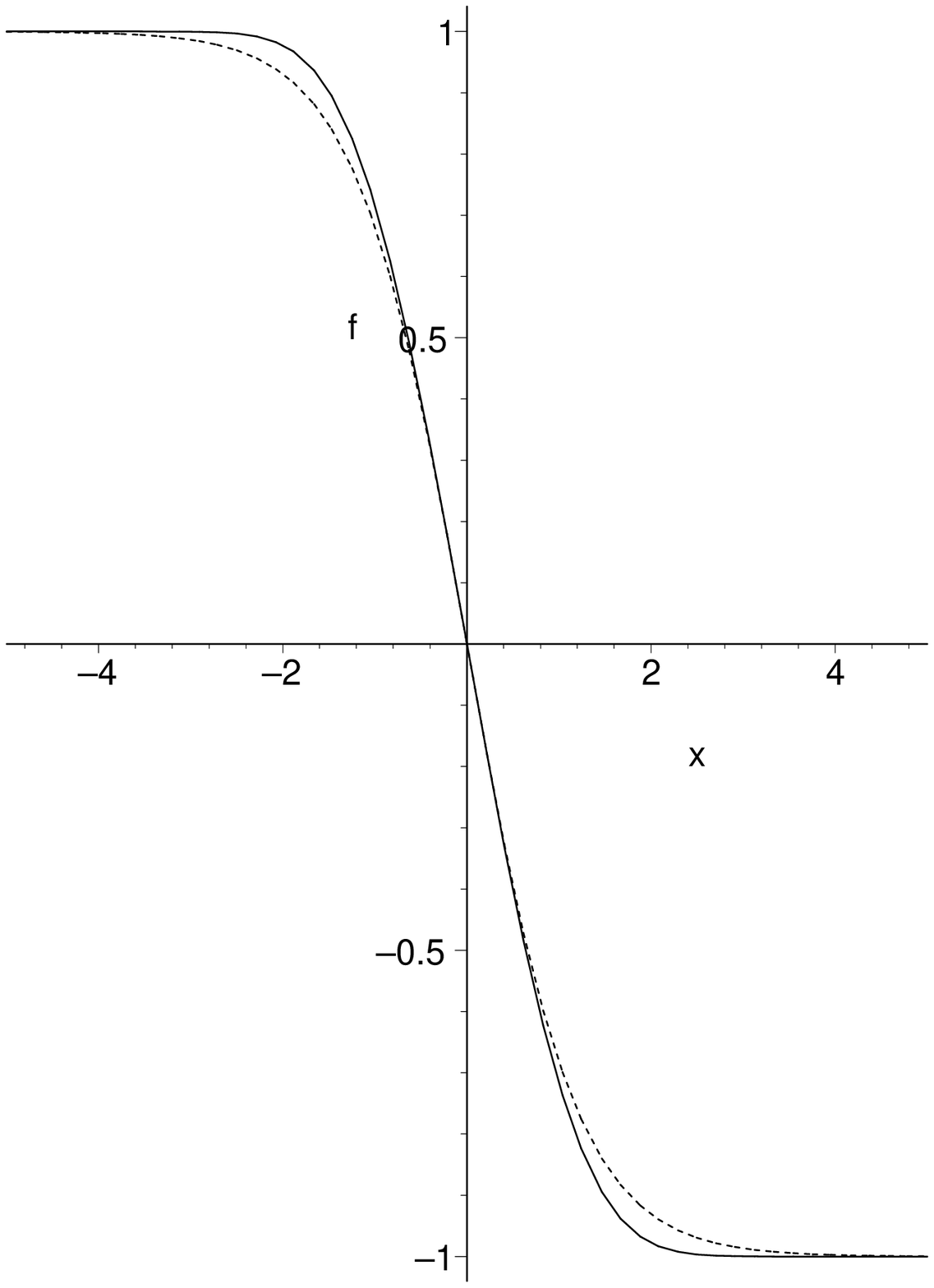}}
\hspace{0.4cm}
\subfigure[Relative error]{\includegraphics[height=3.3cm,width=3.5cm]{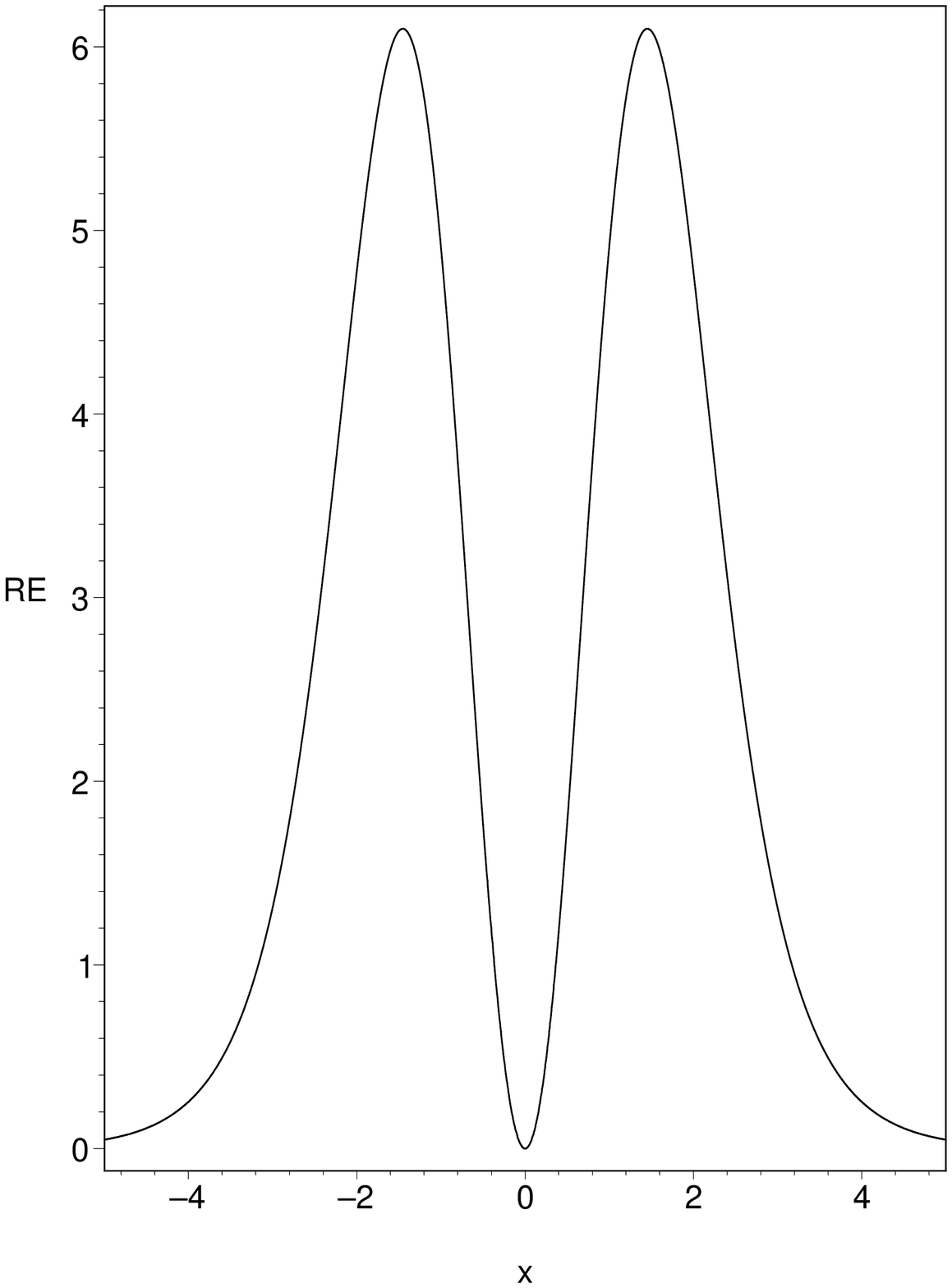}}
\caption{
\label{fig:10}
The drift term $f(x)$ and the approximative one \eqref{tanh} and the relative error (percentage) 
for $\lambda =1,\mu =1$ and $D=1$}
\end{figure}
\begin{figure}
\centering
\includegraphics[height=6.0cm,width=6.0cm]{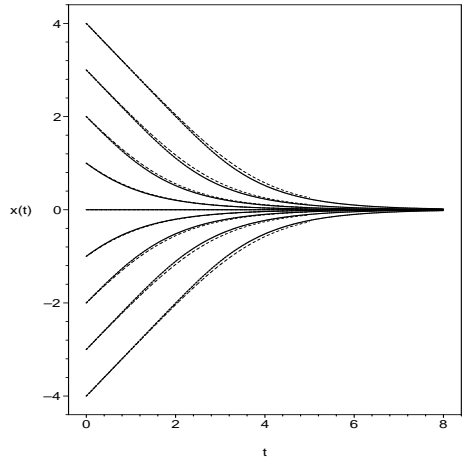}
\caption{\label{fig:11} Trajectories $x(t)$ (solid) and the approximative trajectories $x_a (t)$ 
(dashdotted) $\lambda =1,\mu =1, D=1$ and $-4 \leq x_0 \leq 4$}
\end{figure}
The approximative drift term leads to the following solution for the trajectories
\begin{equation*}
x_a (t)=\frac{\sqrt{\mu\, D}}{s}\, \text{arsinh}\left[\sinh\left(\frac{s}{\sqrt{\mu\, D}}\, x_0\right)\, \text{e}^{s\, t}\right] \quad ,
\end{equation*}
which are plotted in comparison to the numerically calculated trajectories in Fig.~\ref{fig:11}. 
Approximation \eqref{tanh} underestimates the exact drift term that leads to an overestimated absolute 
value of $x(t)$ in the intermediate regime, whereas for the piecewise approximation the drift term is 
overestimated and (the absolute value of) the  deviation between $x(t)$ and $x_a (t)$ in the cross-over 
area is underestimated. A similar situation is observed for the exponential decreasing initial condition. Therefore, 
we skip this part.\\
Concluding our paper we have discussed a simple evolution equation with a long-range memory. Due to the 
permanent coupling to the initial distribution the system offers a stationary solution which depends on the 
special choice of the initial distribution. Then we have demonstrated that this evolution equation is 
fully related to a FPE with non-trivial deterministic forces. As a new feature it results that the 
deterministic part is also characterized by the diffusion constant $D$, with other word the memory 
induces a stochastic behavior also within the deterministic part. As the consequence already the 
deterministic part of the underlying Langevin equation exhibits special trajectories.

\subsection*{Acknowledgments}
The authors (S.~T. and K.~Z.) acknowledge support by the DFG 
(SFB ~418) as well as by DAAD (S.~Tatur).

\newpage

\end{document}